\documentclass[superscriptaddress,reprint,aps,prd,floatfix]{revtex4-2}
\usepackage{color}
\usepackage{amsmath,amssymb}
\usepackage{graphicx}
\usepackage[colorlinks, linkcolor=blue, citecolor=blue, urlcolor=blue, breaklinks=true]{hyperref}
\usepackage{physics}
\usepackage{mathtools} 
\usepackage[title]{appendix}

\usepackage{xr}
\begin{document}
\title{How gravitational fluctuations degrade the high-dimensional spatial entanglement}
\author{Haorong Wu}
\affiliation{Department of Physics and Collaborative Innovation Center for Optoelectronic Semiconductors and Efficient Devices, Xiamen University, Xiamen 361005, China}
\author{Xilong Fan}
\email{xilong.fan@whu.edu.cn}
\affiliation{School of Physics and Technology, Wuhan University, Wuhan 430072, China}
\author{Lixiang Chen}
\email{chenlx@xmu.edu.cn}
\affiliation{Department of Physics and Collaborative Innovation Center for Optoelectronic Semiconductors and Efficient Devices, Xiamen University, Xiamen 361005, China}

\begin{abstract}
Twisted photons carrying orbital angular momentum (OAM) are competent candidates for future interstellar communications. However, the gravitational fluctuations are ubiquitous in spacetime. Thus a fundamental question arises naturally as to how the gravitational fluctuations affect the coherence and the degree of high-dimensional OAM entanglement when twisted photons travel across the textures of curved spacetime. Here, we consider the covariant scalar Helmholtz equations and the Minkowski metric with fluctuations of Gaussian distribution and formulate analytically the equations describing the motion for twisted light in the Laguerre-Gaussian mode space. It is seen that the OAM cannot remain conserved in the presence of gravitational fluctuations. Furthermore, two-photon density matrices are derived for interstellar OAM quantum entanglement distribution, and the degree of entanglement degradation is characterized by purity and negativity. It is revealed that the higher-dimensional OAM entanglement is more susceptible to spacetime fluctuations. We believe that our findings will be of fundamental importance for the future interstellar quantum communications with twisted photons.
\end{abstract}
\maketitle

\section{Introduction}
Multiple degrees of freedom, such as wavelength, polarization, and time-bins, have been utilized to convey information in optical communications \cite{giustina2015significant,marcikic2003long}. Additionally, entanglements of these degrees of freedom are the primary sources of quantum communications. However, since these entanglements may be altered during propagation under the impact of media, their transmission must be investigated to compensate for these modifications. Unlike polarization, which is specified in a two-dimensional space, orbital angular momentum (OAM) of light may assume well-defined values of $l \hbar $ where $l=0, \pm 1, \pm 2, \dots$, that span an infinite-dimensional Hilbert space \cite{allen1992orbital,calvo2006quantum,mair2001entanglement,molina2001management,leach2002measuring,vaziri2002experimental,vaziri2003concentration,allen1999iv,kivshar2001optical,molina2007twisted,yao2011orbital,molina2001propagation}. Due to the fact that information may be encoded in a high-dimensional space, light with OAM is a potential source for future quantum communications \cite{miller2018higher}. However, OAM entanglements are suffered from the adverse effects from the environment. For instance, some investigations have shown the degradation of OAM entanglements as a result of atmospheric turbulence \cite{roux2011infinitesimal,roux2015entanglement,paterson2005atmospheric,roux2013decay,rodenburg2012influence}. As a result, the OAM light transmission distance remains inside the confines of 
a city \cite{peplow2014twisted,krenn2014communication}. Fortunately, in outer space, OAM entanglement will be free from these harmful atmospheric effects. 

In recent years, there were several attempts to connect the photon OAM detection with astronomical observations. Harwit briefly discussed the possible astronomical apparatus for detecting photon OAM as well as their limitations \cite{harwit2003photon}. It was revealed by Tamburini {\it et al.} that it is possible to measure the twisted light for a direct observational demonstration of the existence of rotating black holes, arising from the relativistic effect of their surrounding space and time \cite{tamburini2011twisting}. By extracting the OAM spectra from the radio intensity data collected by the Event Horizon Telescope, the first observational evidence of twisted light from a rotating black hole M87* was presented by Tamburini and coworkers \cite{tamburini2020measurement}. Also, the effects of reduced spatial coherence of astronomical light sources and that of line-of-sight misalignment on the OAM detection were considered carefully by Hetharia {\it et al} \cite{hetharia2014spatial}. 

Photons move not just through a medium in interstellar communications, but also across the textures of spacetime, requiring consideration of general relativity \cite{ursin2007entanglement,yin2017satellite,yin2017satellite2}. Recent years have also witnessed a growing research interest in how coherence and quantum entanglement was affected in curved spacetime \cite{ralph2009quantum,ralph2014entanglement,joshi2018space,xu2019satellite, anastopoulos2013master,lagouvardos2021gravitational, exirifard2021towards,xu2020toy,lamine2002gravitational,kok2003gravitational,wang2006quantum,oniga2016quantum,bassi2017gravitational,korbicz2017information}. To name a few, the event formalism was proposed to solve the closed timelike curves (CTC) problems, although a recent experiment with the quantum satellite {\textit{Micius}} did not support its prediction that time-energy entangled photons would decorrelate after passing through different regions of gravitational potential \cite{ralph2009quantum,ralph2014entanglement,joshi2018space,xu2019satellite}; a master equation for photons was developed by modeling decoherence arising from a bath of stochastic gravitational disturbances \cite{anastopoulos2013master,lagouvardos2021gravitational}; the quantum field theory of light in generally curved spacetime geometry was studied by Exirifard {\it et al.} and the leading corrections by the Riemann curvature to quantum optics are calculated \cite{exirifard2021towards}. Hitherto, to the best of our knowledge, the study for degradation of high-dimensional OAM entanglement due to gravitational fluctuations has not yet been conducted, which, however, is of fundamental importance for future interstellar communications with twisted photons.

Gravitational fluctuations may come from a variety of sources. Classically, they can be a remnant of primordial gravitational waves generated during the early stage of the universe, a remnant of the big bang as the cosmic microwave background, or the sum of the gravitational waves randomly produced by a vast variety of sources in the universe \cite{bassi2017gravitational,christensen2018stochastic}. Also, gravitational fluctuations can arise due to the quantum gravity, which is still in debate, in the form of quantum noise \cite{zhang2021gravitons,parikh2020noise}. Due to the fact that the gravitational fluctuations are ubiquitous in spacetime, it is of fundamental importance to investigate the effect of gravitational fluctuations on the high-dimensional OAM entanglement for future interstellar communications. Here, we formulate analytically the equations describing the motion for twisted light in the Laguerre-Gaussian mode space based on the Minkowski metric and fluctuations. By taking ensemble averages over fluctuations, the evolution of two-photon density matrices for high-dimensional OAM states is derived and the degree of entanglement degradation is measured by calculating its negativity. We reveal that the higher-dimensional OAM entanglement is more susceptible to spacetime fluctuations.

We will operate in position space, i.e., expanding OAM states into Laguerre-Gaussian (LG) modes, ${\rm LG}_{l,p}(\mathbf x)$. Several integrals involving LG modes and other functions will be produced. Due to the fact that LG modes include associated Laguerre polynomials, solving the integrals will be challenging. Fortunately, this issue may be overcome by using the generating-function approach. Unless otherwise specified, geometrized units with $c=G=1$ are used. The signature of metrics will be chosen as $(-,+,+,+)$. The following conventions apply to indices: all Greek indices run in $\{0, 1, 2, 3\}$, whereas Latin indices run in $\{1, 2, 3\}$. The following assumptions and approximations are made for simplicity. First, the beam is assumed to be monochromatic and the paraxial approximation is applied; second, the polarization degrees of freedom will be neglected, since light fields with different polarizations will not couple to each other in the following analysis and polarizations obey the superposition principle; third, the geodesic via which photons propagate will be approximated by a straight line due to the weak magnitudes of fluctuations. Throughout this paper, $x$ or $x^\mu$ stand for the four position vector $(x^0,x^1,x^2,x^3)$, $\mathbf x$ for its spatial components $(x^1,x^2,x^3)$ and $\mathbf x_\perp$ for the spatial transverse components $(x^1,x^2)$.


\section{Light wave in metric fluctuations}
\begin{figure}[ht]
	\centering
	\includegraphics[width=.8\linewidth]{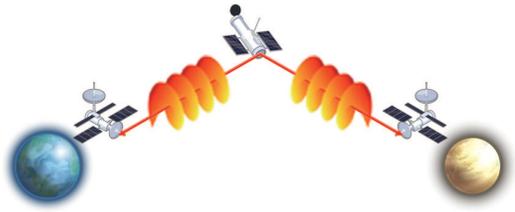}
	\caption{Scheme for interstellar OAM entanglement transmission. The image of satellites and planets are designed by macrovector/Freepik.}
	\label{Fig.lightPath}
\end{figure}

We start by considering two entangled light beams traveling between satellites as in Fig. \ref{Fig.lightPath}. Suppose the metric experienced by them can be expressed as $
g_{\mu\nu}(x)=\eta_{\mu\nu}+\epsilon h_{\mu\nu}(x)
$, where $\epsilon$ is a systematic perturbation parameter which will be set to one in the end, $\eta_{\mu\nu}$ is the Minkowski metric and $h_{\mu\nu}(x)$ is the metric fluctuations. As known, there are a variety of fluctuation sources, e.g., a remnant of primordial gravitational waves generated during the early stage of the universe, gravitational waves randomly produced by a vast variety of sources in the universe, and the quantum gravity effects. Hence, according to the law of large numbers, we can assume the spacetime to be fluctuating around the Minkowski metric, i.e., $\left <h_{\mu\nu}(x)\right >=0$, and isotropic such that its autocorrelation obeys a Gaussian distribution \cite{asprea2021gravitational,asprea2021gravitational1} $
\left < h_{\mu\nu}(x) h_{\mu\nu}(x')\right > =A^2 \delta_{\mu\nu} \exp \left (- \sum_{\mu=0}^3\left (x^\mu-x^{\mu'} \right )^2 /L^2 \right )
$, where $A$ is the fluctuation strength, and $L$ is the correlation length of the fluctuation. Points with distance less than $L$ behave coherently. Later, when we take ensemble averages over Eq. \eqref{eq.integralEOM}, only the spatial fluctuations make an impact on the OAM light, while the temporal parts have no effects. Therefore, the choice of temporal correlation functions will not affect our results of spatial OAM mode distribution. Without loss of generality, we have followed Ref. \cite{bassi2017gravitational} to treat the temporal and spatial components of fluctuations equally, namely, the same correlation length $L$ is assumed. Also, as in \cite{asprea2021gravitational}, we assumed that $h_{\mu\nu}(x)\ll 1$. As a result, only the first-order perturbation is required in our calculation. 

In the scalar theory of light propagation in free space, scalar light field distributions satisfy the Helmholtz equation \cite{cao1998power}, $\eta^{\mu\nu}\psi_{,\mu,\nu}=0$, where commas represent partial derivatives. This equation is already Lorentz invariant and $\psi(x^\mu)$ are scalar fields, whose second-order covariant derivatives are invariant when the order of derivatives change, so according to the comma-goes-to-semicolon rule \cite{carroll2019spacetime,misner1973gravitation}, its counterpart in curved spacetime can be written as $
\Box \psi=g^{\mu\nu}\psi_{;\mu;\nu}=g^{\mu\nu}(\psi_{,\mu,\nu}-\Gamma^{\lambda}_{~\mu\nu}\psi_{,\lambda})=0
$, where the semicolons mean covariant derivatives, $\Gamma^{\lambda}_{~\mu\nu}$ are the Christoffel symbols and we have used the fact that $\psi(x)$ is a scalar field. This equation can also be derived from the action of a scalar field in curved spacetime, $S[\psi]=\int dx^4 \sqrt{-g}g^{\mu\nu}\psi_{;\mu}\psi_{;\nu} /2$, by varying it with respect to $\psi$, where $g$ is the determinant of the metric. We have assumed the light beam propagates along the $x^3$ direction so $\psi(x)=T(\mathbf x) e^{ik(x^0-x^3)}$, where $k$ is the frequency and $T(\mathbf x)$ is the spatial structure mode. For a Gaussian beam, $T(\mathbf x)$ satisfies the paraxial wave equation, whose solutions may be represented as Hermite-Gaussian (HG) modes, Laguerre-Gauss (LG) modes, Ince-Gaussian (IG) modes \cite{exirifard2021towards}, etc. Substituting the fluctuation metric into this equation and keeping terms up to the first order of $\epsilon$, we obtain the variations of the transverse modes given by
\begin{equation}
	\partial_3 T(\mathbf x) \approx\partial_3^{(M)}T(\mathbf x)+\epsilon\partial_3^{(F)}T(\mathbf x), \label{Eq.dT}
\end{equation} 
where $\partial_3^{(M)}T(\mathbf x)=(2ik)^{-1} \nabla_{\perp}^2T(\mathbf x) 
$ is the variations in the Minkowski spacetime, and $\partial_3^{(F)} T(\mathbf x)=(2i)^{-1}k (h_{00}+h_{33}) T(\mathbf x)+(2ik)^{-1}[(h_{33}-h_{11})\partial^2_1T(\mathbf x)+(h_{33}-h_{22})\partial^2_2T(\mathbf x)]$ is the first-order corrections due to the gravitational fluctuations. We will use the superscripts $(M)$ and $(F)$ to distinguish these light field variations. To obtain this result, we have used the paraxial approximation to drop $\partial^2_3T(\mathbf x) $, and terms containing $k^{-2}$ are omitted due to the large magnitude of the frequency $k$. Later, we will find that only terms proportional to $k^{-1}$ have effects.

\section{Equations of motion for OAM states}
We will consider light beams whose transverse components have LG-mode profiles, so $T(\mathbf x)=\text{LG}_{l,p}(\mathbf x_\perp,x^3)$, where $l$ is called the azimuthal index (or topological charge), and $p$ is the radial index. Since we are mostly interested in the entanglement between different azimuthal numbers, we will set the radial number $p$ to zero, implicitly. Hence, $\ket{l, x^3} $ stands for an OAM state with azimuthal number $l$ and $x^3$ indicates that the mode state will be decomposed in the position space at point $x^3$ as $
\ket{ l,x^3}=\int d^2 \mathbf x_\perp {\rm LG}_{l}(\mathbf x_\perp, x^3)\ket{ \mathbf x_\perp }
$.

In the Minkowski spacetime, since the $\text{LG}$ modes are the solutions of the paraxial wave equation, the two OAM indices $l$ and $p$ will be conserved over the propagation. However, in the presence of gravitational fluctuations, the OAM state will evolve according to $
\partial_3 \ket{l,x^3} = \int d^2 \mathbf x_\perp \left [\partial_3^{(M)}{\rm LG}_m(\mathbf x_\perp, x^3)+\partial_3^{(F)}{\rm LG}_m(\mathbf x_\perp, x^3)
\right ] \ket{ \mathbf x_\perp }
$, where the perturbation parameter is set to 1. Obviously an OAM state may spread to an OAM spectrum, i.e., $
\ket{l,x^3} \rightarrow \ket{\tilde l,x^3+\Delta x^3}=\sum_{m} C_m\ket{ m,x^3+\Delta x^3}
$, where $\{ C_m \}$ is a set of constants. A tilde symbol is used to indicate that this is a nonstandard OAM state. Especially, $\left \{\ket{\tilde l,x^3+\Delta x^3} \right \}$ are not generally orthonormal. In other words, the OAM for single-photon states after experiencing the fluctuations cannot remain conserved well. This is because in the Minkowski metric without fluctuations, the spacetime is isotropic and homogeneous. However, the presence of fluctuations will break the spacetime symmetry, thus leading to the non-conservation and even the decoherence of single OAM states.

Suppose, in the future, we would establish a quantum communication channel by distributing two entangled OAM beams as in Fig. \ref{Fig.lightPath}. The density operator for them is given by $
\rho(x^3)=\sum_{l_1,l_2,j_1,j_2} \ket{l_1,x^3 }\ket{l_2,x^3 } \rho_{l_1,l_2,j_1,j_2}(x^3) \bra{ j_1,x^3} \bra{ j_2,x^3} 
$,
where $l_1$, $l_2$, $j_1$, $j_2$ are OAM indices. In a propagation over a small distance $\Delta x^3$, we let the OAM basis evolve and then project them back to the standard OAM basis. In supplemental material, it will be shown that the equations of motion (EOM) for density operator are given by $
\partial_3 \rho_{l_1 l_2,j_1 j_2}=L^{*}_{l_1,m_1}\rho_{m_1 l_2, j_1 j_2}+L^{*}_{l_2,m_2}\rho_{l_1 m_2,j_1 j_2}+L_{j_1,n_1}\rho_{l_1 l_2, n_1 j_2}+L_{j_2,n_2}\rho_{l_1 l_2, j_1 n_2} 
$, where repeated indices imply Einstein summation convention and indices $m_1$, $m_2$, $n_1$, $n_2$ run in all azimuthal numbers. The $L$ symbols are given by $
L_{n,s}=\int d^2 \mathbf x_\perp {\rm LG}_n(\mathbf x_\perp ,x^3)\partial_3^{(F)} {\rm LG}_s^{*} (\mathbf x_\perp ,x^3)
$. Notice that in the absence of any gravitational fluctuations, or in Minkowski spacetime, $\partial_3^{(F)} {\rm LG}_l(\mathbf x_\perp ,x^3)$ identically vanish and all $L$ symbols are zero. This leaves the EOM to be $\partial_3 \rho_{l_1 l_2,j_1 j_2}=0$, i.e., the density matrix will remain unchanged if no gravitational fluctuations affect it. Furthermore, due to the fact that the fluctuation strength is very weak, in the evolution of some mode $\ket{ l }$, the crosstalk between $\ket{ l }$ and other modes are much smaller than $\ket{ l }$ itself. Besides, from the definition of $L_{n,s}$, and in the situation where the correlation length is larger than the beam radius, we can approximately calculate $L_{n,s} \propto \int d^2 \mathbf x_\perp  {\rm LG}_n(\mathbf x_\perp ,x^3) \nabla^2_{\perp}{\rm LG}_s^{*} (\mathbf x_\perp ,x^3)$, which is nonzero only when $n=s$. Therefore, we will neglect those crosstalk terms. The EOM will reduce to $
\partial_3 \rho_{l_1 l_2,j_1 j_2}=\left (L^{*}_{l_1,l_1} +L^{*}_{l_2,l_2} +L_{j_1,j_1} +L_{j_2,j_2} \right )\rho_{l_1 l_2, j_1 j_2}
$, where the Einstein summation convention is suppressed. By substituting the expressions of $L$ symbols, one may verify that terms proportional to $k$ cancel each other.

\begin{figure}[ht]
	\centering
	\includegraphics[width=.8\linewidth]{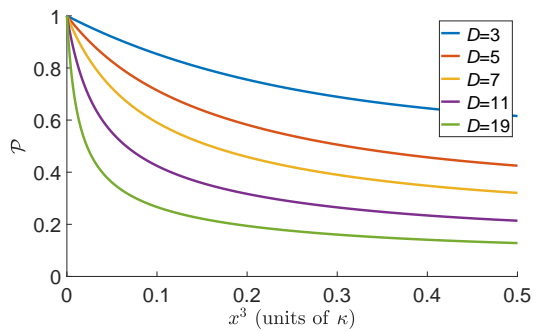}
	\caption{Purities for different dimensional entanglements. }
	\label{Fig.Purity}
\end{figure}

\begin{figure}[ht]
	\centering
	\includegraphics[width=.8\linewidth]{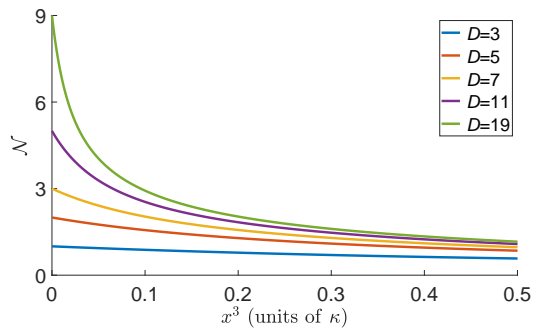}
	\caption{Negativities for different dimensional entanglements.}
	\label{Fig.Negativity}
\end{figure}

Now let the beams propagate from distance $x^3$ to $x^3+\Delta x^3$. Normally, the gravitational fluctuation is very weak, and the beam has to propagate a great distance before its density operator changes significantly. Therefore, we may choose $\Delta x^3 $ to be much greater than $L$, the correlation length of the fluctuation. By invoking the Dyson series to the second-order, and taking ensemble average, one may have the density operator at $(x^3+\Delta x^3)$ as
\begin{align}
	&~~~~\rho_{l_1 l_2,j_1 j_2}(x^3+\Delta x^3) =\rho_{l_1 l_2,j_1 j_2}(x^3)\bigg [ 1+\int_{x^3}^{x^3+\Delta x^3}dx'^3 \nonumber \\& \cross \int_{x^3}^{x'^3}dx''^3
	\bigg  < 
	\left (L^{*}_{l_1,l_1}(x'^3)+L^{*}_{l_2,l_2}(x'^3) +L_{j_1,j_1}(x'^3)\right .\nonumber \\& \left . +L_{j_2,j_2}(x'^3) \right )  
	\left (L^{*}_{l_1,l_1}(x''^3) +L^{*}_{l_2,l_2}(x''^3) +L_{j_1,j_1}(x''^3) \right .\nonumber \\& \left . +L_{j_2,j_2}(x''^3) \right )		
	\bigg >\bigg ], \label{eq.integralEOM}
\end{align}
where $\left < \cdot \right >$ means ensemble average. Note that if one takes ensemble averages over crosstalk terms $L_{l,j}(x'_3)L_{m,n}(x''_3)$, the numerical calculations will show that $\left <L_{l,j}(x'_3)L_{m,n}(x''_3)\right > \ll \left <L_{l,l}(x'_3)L_{m,m}(x''_3)\right > $ if $l\ne j$ or $m \ne n$, so the crosstalk terms can be safely neglected. After the ensemble averages are taken, we may utilize the generating-function method to find derivatives for each element. In astrophysics and cosmology, fluctuations occur over large distances \cite{asprea2021gravitational}, so we consider the situation where the correlation length is larger than the beam radius, i.e., $L\gg w(z)$. In this case, Equation \eqref{eq.integralEOM} can be approximated by $
\partial_3 \rho_{l_1 l_2,j_1 j_2}(x^3)=-C_{l_1 l_2,j_1 j_2} \rho_{l_1 l_2,j_1 j_2}(x^3)/\kappa
$, where $\kappa= 2 k^2 w_0^4/(3 LA^2)$ is a characteristic length and $C_{l_1 l_2,j_1 j_2}= (\vert l_1\vert-\vert j_1\vert)^2+(\vert l_2\vert- \vert j_2\vert)^2$. Therefore, at any distance $x^3$, we have \begin{equation}
	\rho_{l_1 l_2,j_1 j_2}(x^3)= \rho_{l_1 l_2,j_1 j_2}(0) \exp \left (-C_{l_1 l_2,j_1 j_2} x^3/\kappa  \right ).
\end{equation} Obviously, we can see a trivial case that $C_{l_1 l_2,j_1 j_2}=0$ if $\vert l_1\vert=\vert j_1\vert$ and $\vert l_2\vert=\vert j_2\vert$ and hence that the trace of density operator is not altered along the propagation. This property can also be seen from the definition of $L_{n,s}$, from which we have $L_{l,l}^*=-L_{l,l}$, and $L_{-l,-l}=L_{l,l}$. Hence when $|l_1|=|j_1|$ and $|l_2|=|j_2|$, the fluctuation-induced correction $L_{l_1,l_1}^*$ will be canceled by another correction $L_{j_1,j_1}$; similarly, $L_{l_2,l_2}^*$ will be canceled by $L_{j_2,j_2}$. Therefore, these elements will be shelled from the fluctuations, and be protected from those adverse effects.

\section{High-dimensional entanglement in gravitational fluctuation}
We will consider a high-dimensional entangled state $
\ket{ \psi_{\text {OAM}}(0)}=\sum_{m=-M}^M\ket{ m }\ket{ -m}/\sqrt{D}
$, where $D=2M+1$ is the dimension of the system. We use the purity, defined by $\mathcal P={\rm tr}(\rho^2)$ \cite{chatterjee2021density,zurek1993coherent,isar1999purity}, to measure the coherence of the system. It can be shown to be $
\mathcal P (x^3)= \sum_{l,j} \exp \left ( - 2C_{l,-l,j,-j}    x^3  /\kappa  \right )
/D^2$. We can see that the coherence will drop exponentially as $x^3$ grows. Also, a higher-dimensional system will lose coherence more rapidly. For measuring entanglement, we use the negativity $\mathcal N$ \cite{horodecki2009quantum,rungta2001universal,anaya2019neumann,mintert2005measures,guhne2009entanglement}, defined by $
{\mathcal N}(\rho)=-\sum_{\lambda_k <0} \lambda_k
$
where $\lambda_k$ are eigenvalues of the partial transpose of $\rho$, e.g., with respect to the second particle, $
\rho^{PT}=\sum_{l_1, l_2,j_1,j_2}\ket{l_1 }\ket{j_2 } \rho_{l_1 l_2,j_1 j_2}\bra{ j_1}\bra{ l_2}
$. For our system, the partial transpose density, with respect to the second particle, is $
\rho^{\text{PT}}_{\text{OAM}}= \sum_{l,j} \exp \left (-C_{l,-l,j,-j}  x^3 /\kappa\right ) \ket{l} \ket{ -j } \bra{ j }\bra{ -l }/D
$. We could numerically solve its eigenvalues to compute its negativity.

\begin{figure}[tbhp]
	\centering
	\includegraphics[width=.8\linewidth]{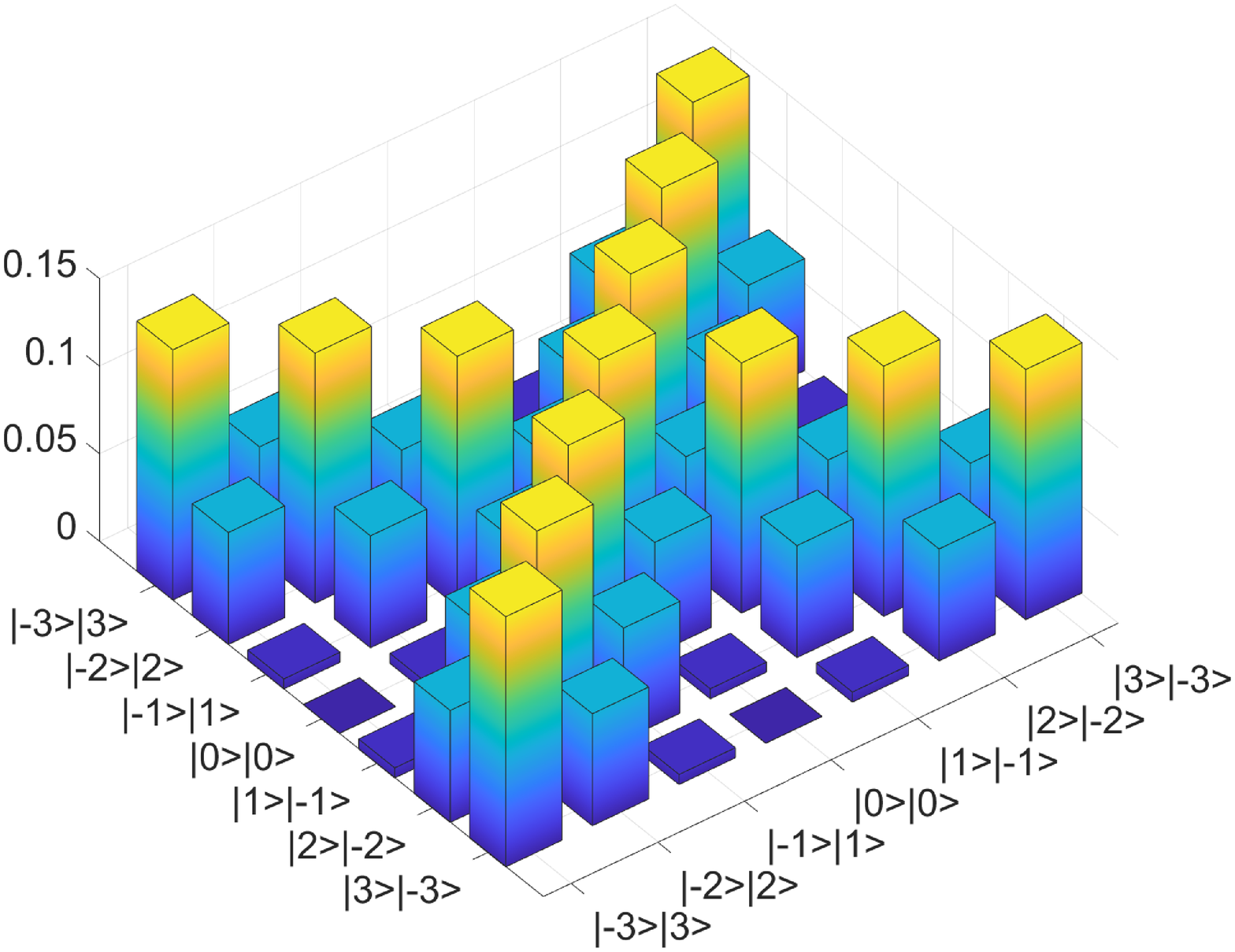}
	\caption{Density matrix for $D=7$ at the decay distance $x^3_{\text{D}}$. Only nonzero elements $\ket{ l }\ket{ -l }\bra{ j}\bra{ -j }$ with $l,~j=-M,~-M+1,~\dots,~M$, are shown.}
	\label{Fig.densitymatrix}
\end{figure}

We vary $D$ for different dimensions and plot them in Figs. \ref{Fig.Purity} and \ref{Fig.Negativity}. These figures show that the coherence and entanglement of a high-dimensional system will decline exponentially as their propagation distances increase. We define the decay distance as the length by which the negativity of the system drops to $1/e$ of the initial value, i.e., $
\mathcal N(x^3_{\text{D}})=\mathcal N(0)/e
$. Numerically, we find that the decay distances are $1.48\kappa $, $0.64\kappa $, $0.40\kappa $, $0.20\kappa $, $0.08\kappa $ for $D=3,~5,~7,~11,~19$, respectively, which shows that higher-dimensional system will suffer more severe degradation. Moreover, from the property of $C_{l_1 l_2,j_1 j_2}$, some specific elements in density operators are free from fluctuations. Entanglements stored in these safety islands will be preserved as depicted in Fig. \ref{Fig.densitymatrix} for $D=7$. Note in this figure, only nonzero elements are shown. Therefore, the entanglements of those systems will never totally vanish. This fact could be an advantage of OAM entanglements. For higher-dimensional systems, even if there are more safety islands, considerably more elements will be affected by fluctuations. Higher-dimensional systems, in other words, have more complex structures that are more susceptible to fluctuations.

The degradation strength is characterized by the characteristic length $\kappa= 2 k^2 w_0^4/(3 LA^2)$. Therefore, if we would like to increase the stability of entanglements to transfer information, we could increase the waist radius $w_0$ or the frequency $k$, or lower the dimension of the system. On the contrary, a stronger degradation strength can be achieved by doing the opposite, in order to experimentally quantify the parameters, $A$ and $L$, of fluctuations. Even so, this kind of experiments are out of current technologies.

\section{Conclusion}
We have investigated the covariant scalar wave equations in gravitational fluctuations and constructed the equations of motion for light with OAM. The evolution of OAM states in gravitational fluctuations is achieved by taking its ensemble average. The high-dimensional entanglements are explored and characterized by purities and negativities. Our estimations indicate that the system will undergo some decoherence, leading to the non-conservation of OAM for single-photon states and to the degradation of OAM entanglement for two-photon states. The parameters of gravitational fluctuations, $A$ and $L$, are still unknown to us. We could use OAM entanglements to measure them, or protect these entanglements from them by varying the parameters of the light beams. Recently, quantum entanglement distribution, with the polarization state of light, has been reported over a distance of 1200 kilometers based on a satellite, towards a global scale \cite{yin2017satellite}. In the future, if the high-dimensional OAM entanglement is utilized to establish interstellar quantum communications, our result may serve as a model for how it will evolve as it travels across the spacetime. Especially, higher dimensional entanglements have more information capacities, while being more susceptible to fluctuations, and this balance should be taken into account.

\section*{Acknowledgments}
We would like to thank Filippus S. Roux for insightful discussions. This work is supported by the National Natural Science Foundation of China (12034016, 11922303), the Fundamental Research Funds for the Central Universities at Xiamen University (20720190057, 20720200074), the Natural Science Foundation of Fujian Province of China for Distinguished Young Scientists (2015J06002), and the program for New Century Excellent Talents in University of China (NCET-13-0495).

\onecolumngrid
\begin{appendices}
\setcounter{equation}{0}
\renewcommand\theequation{A.\arabic{equation}}

\section{Derivation of EOM for density operator}
In a small distance, let the OAM basis change while $\rho_{l_1 l_2,j_1 j_2}(z)$ are unaltered, such that \begin{align}
	\rho(z+\Delta z)&=\sum_{l_1,l_2,j_1,j_2} \left |\tilde l_1,z+\Delta z \right >\left |\tilde l_2,z+\Delta z \right > \rho_{l_1l_2,j_1j_2}(z) \left < \tilde j_1,z+\Delta z\right |\left < \tilde j_2,z+\Delta z\right | \nonumber \\ 
	&=\sum_{l_1,l_2,j_1,j_2}\int d^2\mathbf x_{1\perp} d^2\mathbf x_{2\perp}   d^2\mathbf x'_{1\perp}d^2\mathbf x'_{2\perp}\tilde {{\rm LG}}_{l_1}(\mathbf x_{1\perp},z+\Delta z)\tilde {{\rm LG}}_{l_2}(\mathbf x_{2\perp},z+\Delta z)  \tilde {{\rm LG}}^*_{j_1}(\mathbf x'_{1\perp},z+\Delta z)\tilde {{\rm LG}}^*_{j_2}(\mathbf x'_{2\perp},z+\Delta z)\nonumber\\ &~~~~\times\left |\mathbf x_{1\perp} \right >\left |\mathbf x_{2\perp} \right > \rho_{l_1l_2,j_1j_2}(z)\left <\mathbf x'_{1\perp}\right | \left <\mathbf x'_{2\perp}\right |  \nonumber \\ 
	&=\sum_{l_1,l_2,j_1,j_2}\int d^2\mathbf x_{1\perp} d^2\mathbf x_{2\perp}   d^2\mathbf x'_{1\perp}d^2\mathbf x'_{2\perp}
	   [{\rm LG}_{l_1}(\mathbf x_{1\perp},z) +\Delta z \partial_3{\rm LG}_{l_1}(\mathbf x_{1\perp},z) ]
	   [{\rm LG}_{l_2}(\mathbf x_{2\perp},z) +\Delta z \partial_3{\rm LG}_{l_2}(\mathbf x_{2\perp},z) ]\nonumber\\ &~~~~\times
	    [{\rm LG}^*_{j_1}(\mathbf x'_{1\perp},z) +\Delta z \partial_3{\rm LG}^*_{j_1}(\mathbf x'_{1\perp},z) ][{\rm LG}^*_{j_2}(\mathbf x'_{2\perp},z) +\Delta z \partial_3{\rm LG}^*_{j_2}(\mathbf x'_{2\perp},z) ] \left |\mathbf x_{1\perp} \right >\left |\mathbf x_{2\perp} \right > \rho_{l_1l_2,j_1j_2}(z)\left <\mathbf x'_{1\perp}\right | \left <\mathbf x'_{2\perp}\right |,
\end{align}
where $\tilde {{\rm LG}}_l(\mathbf x_\perp,z+\Delta z)$ is the position-space decomposition of the nonstandard OAM states $\{\left |\tilde l,z+\Delta z\right >\}$, and since the beam has experienced a small propagation, it can be expanded in terms of ${{\rm LG}}_l(\mathbf x_\perp,z)$ as \begin{equation}
	\tilde {{\rm LG}}_l(\mathbf x_\perp,z+\Delta z)={\rm LG}_l(\mathbf x_\perp,z) +\Delta z \partial_3{\rm LG}_l(\mathbf x_\perp,z).
\end{equation}Since $\left |\tilde l,z+\Delta z \right >$ is not a standard OAM state, the elements will be extracted by using $ \left < l_1,z+\Delta z\right | \left < l_2,z+\Delta z\right |\rho(z+\Delta z) \left |j_1,z+\Delta z \right > \left |j_2,z+\Delta z \right >$. Therefore, \begin{align}
	\rho(z+\Delta z)_{l_1,l_2,j_1,j_2}&=\left < l_1,z+\Delta z\right | \left < l_2,z+\Delta z\right |\rho(z+\Delta z) \left |j_1,z+\Delta z \right > \left |j_2,z+\Delta z \right > \nonumber \\ 
	&=\int d^2\mathbf x_{3\perp} d^2\mathbf x_{4\perp} d^2\mathbf x'_{3\perp} d^2\mathbf x'_{4\perp} {\rm LG}^*_{l_1}(\mathbf x_{3\perp},z+\Delta z){\rm LG}^*_{l_2}(\mathbf x_{4\perp},z+\Delta z){\rm LG}_{j_1}(\mathbf x'_{3\perp},z+\Delta z){\rm LG}_{j_2}(\mathbf x'_{4\perp},z+\Delta z)\nonumber\\ &~~~~\times\left <\mathbf x_{3\perp} \right |\left <\mathbf x_{4\perp} \right | \rho(z+\Delta z) \left |\mathbf x'_{3\perp}\right >\left |\mathbf x'_{4\perp}\right > \nonumber \\ 
	&=\sum_{m_1,m_2,n_1,n_2}\int d^2\mathbf x_{1\perp} d^2\mathbf x_{2\perp} d^2\mathbf x_{3\perp} d^2\mathbf x_{4\perp}  d^2\mathbf x'_{1\perp}d^2\mathbf x'_{2\perp} d^2\mathbf x'_{3\perp} d^2\mathbf x'_{4\perp}     	\nonumber\\ &~~~~\times[{\rm LG}^*_{l_1}(\mathbf x_{3\perp},z) +\Delta z \partial^{(M)}_3{\rm LG}^*_{l_1}(\mathbf x_{3\perp},z) ][{\rm LG}^*_{l_2}(\mathbf x_{4\perp},z) +\Delta z \partial^{(M)}_3{\rm LG}^*_{l_2}(\mathbf x_{4\perp},z) ]
	\nonumber\\ &~~~~\times[{\rm LG}_{j_1}(\mathbf x'_{3\perp},z) +\Delta z \partial^{(M)}_3{\rm LG}_{j_1}(\mathbf x'_{3\perp},z) ] [{\rm LG}_{j_2}(\mathbf x'_{4\perp},z) +\Delta z \partial^{(M)}_3{\rm LG}_{j_2}(\mathbf x'_{4\perp},z) ]\nonumber\\ &~~~~\times
	 [{\rm LG}_{m_1}(\mathbf x_{1\perp},z) +\Delta z \partial_3{\rm LG}_{m_1}(\mathbf x_{1\perp},z) ]
	[{\rm LG}_{m_2}(\mathbf x_{2\perp},z) +\Delta z \partial_3{\rm LG}_{m_2}(\mathbf x_{2\perp},z) ]\nonumber\\ &~~~~\times
	[{\rm LG}^*_{n_1}(\mathbf x'_{1\perp},z) +\Delta z \partial_3{\rm LG}^*_{n_1}(\mathbf x'_{1\perp},z) ][{\rm LG}^*_{n_2}(\mathbf x'_{2\perp},z) +\Delta z \partial_3{\rm LG}^*_{n_2}(\mathbf x'_{2\perp},z) ]
	\nonumber\\ &~~~~\times \rho_{m_1m_2,a_1a_2}(z) \left < \mathbf x_{3\perp} \right | \left . \mathbf x_{1\perp}\right > \left < \mathbf x_{2\perp} \right | \left . \mathbf x_{4\perp} \right >
	\left < \mathbf x'_{3\perp} \right | \left . \mathbf x'_{1\perp}\right > \left < \mathbf x'_{2\perp} \right | \left . \mathbf x'_{4\perp} \right >.\label{eq:rhomn}
\end{align} Then we expand those products and only keep terms up to first order of $\Delta z$ and use Eq. \eqref{M-Eq.dT}. Also, the orthogonality of LG functions states that \begin{equation}
	\int d^2 \mathbf x_{\perp} {\rm LG}_m(\mathbf x_{\perp},z){\rm LG}^*_n(\mathbf x_{\perp},z) =\delta_{mn},
\end{equation}
and by applying the two-dimensional Gauss's theorem, \begin{align}
	&~~\int d^2 \mathbf x_{\perp} [ \partial_3^{(M)} {\rm LG}_m(\mathbf x_{\perp},z){\rm LG}^*_n(\mathbf x_{\perp},z)+{\rm LG}_m(\mathbf x_{\perp},z)\partial_3^{(M)}{\rm LG}^*_n(\mathbf x_{\perp},z)]\nonumber \\ 
	&=\frac 1 {2ik}\int d^2 \mathbf x_{\perp} \nabla_T \cdot [ \nabla_T {\rm LG}_m(\mathbf x_{\perp},z){\rm LG}^*_n(\mathbf x_{\perp},z)-{\rm LG}_m(\mathbf x_{\perp},z)\nabla_T{\rm LG}^*_n(\mathbf x_{\perp},z)]\nonumber \\ &=\frac 1 {2ik}\int_{S_{\infty}} [ \nabla_T {\rm LG}_m(\mathbf x_{\perp},z){\rm LG}^*_n(\mathbf x_{\perp},z)-{\rm LG}_m(\mathbf x_{\perp},z)\nabla_T{\rm LG}^*_n(\mathbf x_{\perp},z)]\cdot \hat {\mathbf n} ds\nonumber \\ &=0,
\end{align}
where $S_{\infty}$ is the boundary surface in infinity. Now Equation \eqref{eq:rhomn} can be simplified into\begin{align}
	&~~~~\rho(z+\Delta z)_{l_1,l_2,j_1,j_2} \nonumber \\&=\rho(z)_{l_1,l_2,j_1,j_2}+\Delta z \left [\sum_{m_1}\rho_{m_1l_2,j_1j_2}(z) \int d^2 \mathbf x_{\perp} {\rm LG}^*_{l_1}(\mathbf x_{\perp},z)\partial_3^{(F)} {\rm LG}_{m_1}(\mathbf x_{\perp},z)+\sum_{m_2}\rho_{l_1m_2,j_1j_2}(z) \int d^2 \mathbf x_{\perp} {\rm LG}^*_{l_2}(\mathbf x_{\perp},z)\partial_3^{(F)} {\rm LG}_{m_2}(\mathbf x_{\perp},z)\right . \nonumber\\ 
	&~~~~\left .+\sum_{n_1}\rho_{l_1l_2,n_1j_2}(z) \int d^2 \mathbf x_{\perp} {\rm LG}_{j_1}(\mathbf x_{\perp},z)\partial_3^{(F)} {\rm LG}^*_{n_1}(\mathbf x_{\perp},z)]+\sum_{n_2}\rho_{l_1l_2,j_1n_2}(z) \int d^2 \mathbf x_{\perp} {\rm LG}_{j_2}(\mathbf x_{\perp},z)\partial_3^{(F)} {\rm LG}^*_{n_2}(\mathbf x_{\perp},z)\right ].
\end{align} Therefore, the equations of motion for density operator are given by 
\begin{equation}
	\partial_3 \rho_{l_1 l_2,j_1 j_2}=L^{*}_{l_1,m_1}\rho_{m_1 l_2, j_1 j_2}+L^{*}_{l_2,m_2}\rho_{l_1 m_2,j_1 j_2}+L_{j_1,n_1}\rho_{l_1 l_2, n_1 j_2}+L_{j_2,n_2}\rho_{l_1 l_2, j_1 n_2}  ,\label{eq:mr}
\end{equation}
where Einstein summation convention is used and $m_1$, $m_2$, $n_1$, $n_2$ run in all azimuthal numbers. The $L$ symbols are given by \begin{equation}
L_{n,s}=\int d^2 \mathbf x_\perp {\rm LG}_n(\mathbf x_\perp ,z)\partial_3^{(F)} {\rm LG}_s^{*} (\mathbf x_\perp ,z). \label{eq:Al}
\end{equation}
	
\setcounter{equation}{0}
\renewcommand\theequation{B.\arabic{equation}}
\section{LG modes and the generating-function method}
The Laguerre-Gaussian (LG) modes which satisfy the paraxial wave equation, \begin{equation}
	-2ik\frac {\partial T}{\partial z}+\frac {\partial^2 T}{\partial r^2}+\frac 1 r \frac {\partial T}{ \partial r}+\frac 1 {r^2}\frac {\partial^2 T}{\partial \phi^2}=0,
\end{equation}
in cylindrical coordinates, are given by \begin{equation}
	{\rm LG}_{l,p}(x,y,z)=\mathcal N \frac 1 {w(z)}(\frac {\sqrt 2 r}{w(z)})^{|l|}{\rm L}^{|l|}_{p}\left ( \frac{2r^2}{w^2(z)}\right )\exp \left (-\frac {r^2}{w^2(z)}+il\phi+i \frac {k r^2}{2 R(z)}+i \Phi(z) \right ),
\end{equation}
where $\mathcal N=\sqrt{\frac{2 p!}{\pi (|l|+p)!}}$ is the normalization constant, $r=\sqrt {x^2+y^2}$, $\phi=\arctan (y/x)$, $w(z)=w_0 \sqrt{1+(z/z_R)^2}$ with $w_0$ being the waist radius and $z_R=kw^2_0/2$ being the Rayleigh range, $R(z)=z[1+(z_R/z)^2]$ is the curvature radius of the wavefronts, $\Phi(z)=-(|l|+2p+1)\arctan (z/z_R)$ is the Gouy phase, $k$ is the wave number, and ${\rm L}^{|l|}_{p}(x)$ is the associated Laguerre polynomials. The radial index $p=0, 1, 2, \dots$ indicates the number of radial nodes of the mode, while the azimuthal index $l=0, \pm 1, \pm 2, \cdots$ corresponds to the topological charge. It can be shown that the azimuthal index $l$ represents the amount of orbital angular momentum carried by every photon in that mode, so the LG modes are the decomposition of OAM basis in position space. 

Since the LG modes contain some associated Laguerre polynomials, it would be difficult to integrate them with themselves or other functions. However, those integrals can be calculated with corresponding generating functions, and then generate the correct result. The generating functions for LG modes are given by \begin{equation}
	{\mathcal G}_{{\rm LG}(l,p)}(a,b,c)=\frac 1 {\Omega(z,c)} \exp [\frac {(x+iy)a w_0+(x-iy)b w_0-(1+c)(x^2+y^2)}{w_0^2\Omega(z,c)}],
\end{equation} where \begin{equation}
\Omega(z,c)=1-c+i\frac{z}{z_R}+ic\frac{z}{z_R},
\end{equation} and $a$, $b$, $c$ are new variables. The generating procedure is given by \begin{equation}
{\rm LG}_{l,p}(x,y,z) =
	\frac{\mathcal N (\sqrt 2)^{|l|}}{w_0\cdot p!} ( \partial^p_c \partial^{|l|}_{a|b} 	{\mathcal G}_{{\rm LG}(l,p)}(a,b,c))  |_{a,b,c=0} ,
\end{equation}
where $\partial^{|l|}_{a|b} $ means taking partial derivative with respect to $a$ if $l >0$, with respect to $b$ if $l<0$, and no partial derivative is taken if $l=0$. Now that the associated Laguerre polynomials are replaced by a exponential function, the integral can be easily calculated. For example, for the integral Eq. \eqref{eq:Al}, the generating functions for LG modes can be substituted in it yielding the generating functions for $L_{n,s}$, \begin{equation}
	{\mathcal G}_{L(n,s)}=\int d^2 \mathbf x_{\perp} {\mathcal G}_{{\rm LG}(n)}(\mathbf x_{\perp},z)\partial_3^{(F)} {\mathcal G}_{{\rm LG}(s)}^{*} (\mathbf x_{\perp},z).  \label{eq:gl}
\end{equation} Then using the generating procedure, we have \begin{equation}
L_{n,s}=\frac{\mathcal N_n (\sqrt 2)^{|l_n|}}{w_0\cdot p_n!} \frac{\mathcal N_s (\sqrt 2)^{|l_s|}}{w_0\cdot p_s!} ( \partial^{p_n}_{c_n} \partial^{p_s}_{c_s} \partial^{|l_n|}_{a_n|b_n}  \partial^{|l_s|}_{a_s|b_s}	{\mathcal G}_{L(n,s)}(x,y,z))  |_{a_n,b_n,c_n,a_s,b_s,c_s=0}.  \label{eq:glsn}
\end{equation}

\end{appendices}
\twocolumngrid
\bibliography{myBib}
\end{document}